\documentstyle[12pt,world_sci]{article}
\begin{document}

\def\Re{{\cal R \mskip-4mu \lower.1ex \hbox{\it e}\,}}
\def\Im{{\cal I \mskip-5mu \lower.1ex \hbox{\it m}\,}}
\def\ie{{\it i.e.}}
\def\eg{{\it e.g.}}
\def\etc{{\it etc}}
\def\etal{{\it et al.}}
\def\ibid{{\it ibid}.}
\def\tev{\,{\rm TeV}}
\def\gev{\,{\rm GeV}}
\def\mev{\,{\rm MeV}}
\def\to{\rightarrow}
\def\slash{\not\!}
\def\mh{\ifmmode m\sbl H \else $m\sbl H$\fi}
\def\mch{\ifmmode m_{H^\pm} \else $m_{H^\pm}$\fi}
\def\mt{\ifmmode m_t\else $m_t$\fi}
\def\mc{\ifmmode m_c\else $m_c$\fi}
\def\mz{\ifmmode M_Z\else $M_Z$\fi}
\def\mw{\ifmmode M_W\else $M_W$\fi}
\def\mws{\ifmmode M_W^2 \else $M_W^2$\fi}
\def\mhs{\ifmmode m_H^2 \else $m_H^2$\fi}
\def\mzs{\ifmmode M_Z^2 \else $M_Z^2$\fi}
\def\mts{\ifmmode m_t^2 \else $m_t^2$\fi}
\def\mcs{\ifmmode m_c^2 \else $m_c^2$\fi}
\def\mchs{\ifmmode m_{H^\pm}^2 \else $m_{H^\pm}^2$\fi}
\def\ztwo{\ifmmode Z_2\else $Z_2$\fi}
\def\zone{\ifmmode Z_1\else $Z_1$\fi}
\def\mtwo{\ifmmode M_2\else $M_2$\fi}
\def\mone{\ifmmode M_1\else $M_1$\fi}
\def\tb{\ifmmode \tan\beta \else $\tan\beta$\fi}
\def\xw{\ifmmode x\sub w\else $x\sub w$\fi}
\def\ch{\ifmmode H^\pm \else $H^\pm$\fi}
\def\lum{\ifmmode {\cal L}\else ${\cal L}$\fi}
\def\inpb{\ifmmode {\rm pb}^{-1}\else ${\rm pb}^{-1}$\fi}
\def\infb{\ifmmode {\rm fb}^{-1}\else ${\rm fb}^{-1}$\fi}
\def\epem{\ifmmode e^+e^-\else $e^+e^-$\fi}
\def\lplm{\ifmmode \ell^+\ell^-\else $\ell^+\ell^-$\fi}
\def\ppb{\ifmmode \bar pp\else $\bar pp$\fi}
\def\subw{_{\rm w}}
\def\half{\textstyle{{1\over 2}}}
\def\elli{\ell^{i}}
\def\ellj{\ell^{j}}
\def\ellk{\ell^{k}}
\newskip\zatskip \zatskip=0pt plus0pt minus0pt
\def\matth{\mathsurround=0pt}
\def\lsim{\mathrel{\mathpalette\atversim<}}
\def\gsim{\mathrel{\mathpalette\atversim>}}
\def\atversim#1#2{\lower0.7ex\vbox{\baselineskip\zatskip\lineskip\zatskip
  \lineskiplimit 0pt\ialign{$\matth#1\hfil##\hfil$\crcr#2\crcr\sim\crcr}}}
\def\undertext#1{$\underline{\smash{\vphantom{y}\hbox{#1}}}$}

\pagestyle{empty}
\setlength{\baselineskip}{2.6ex}
\rightline{\vbox{\halign{&#\hfil\cr
&ANL-HEP-CP-93-51\cr
&August 1993\cr}}}

\title{{\bf USING `INVISIBLE' DECAY MODES AS PROBES OF $Z'$ COUPLINGS}}
\author{J.L.~HEWETT and T.G.~RIZZO\\
\vspace{0.3cm}
{\em High Energy Physics Division, Argonne National Laboratory,\\
Argonne, IL 60439, USA}}
\maketitle

\begin{center}
\parbox{13.0cm}
{\begin{center} ABSTRACT \end{center}
{\small\hspace*{0.1cm}

We explore the possibility that $Z'$ couplings can be probed using decay modes
which involve neutrinos once Standard Model backgrounds are directly
determined by the data itself. For some models, sufficient statistics are
available at either the SSC or LHC to render these modes useful for coupling
determinations, provided the mass of the $Z'$ is not much larger than 1 TeV,
if we assume other new physics background sources are absent.}}

\end{center}

If a new neutral gauge boson, $Z'$, is discovered at the SSC or LHC, an
immediate goal will be the determination of its couplings to fermion pairs.
Knowledge of these couplings will allow us to ascertain the extended
electroweak model from which the $Z'$ has originated. At an \epem\ collider,
such as LEP, the SLC, or the NLC, this is a relatively straightforward
procedure that has already been successfully employed for the Standard Model
(SM) $Z$.  Performing a similar task in the hadron supercollider environment
is far from easy and has prompted much activity in the theoretical and
detector development arenas during the
past three years{\cite {1}}. All of the techniques suggested in Ref.~1 are
somewhat limited in scope so that new tools are continuously being sought to
aid in the determination of $Z'$ couplings at hadron supercolliders.
In this report we outline some preliminary results from an analysis of
$Z'$ decay modes which involve neutrinos. As is always the case, the greatest
difficulty in the use of missing $E_t$ or $p_t$ final states is the size of
the SM backgrounds. In the two cases examined here, however, we may be able
to employ the data itself to precisely determine these backgrounds. To be as
specific as possible, we limit our discussion to only a few extended gauge
models: ($i$) the Left-Right Symmetric Model (LRM){\cite {2}}, ($ii$) the
Alternative Left-Right Model (ALRM){\cite{3}}, ($iii$) a model where the $Z'$
couples exactly like the SM $Z$ (SSM), and ($iv$) the $E_6$-inspired Effective
Rank-5 Model (ER5M){\cite {4}}. In the latter example, the model contains an
additional parameter, $\theta$, which lies in the range $-90^\circ<\theta<
90^\circ$.  Specific choices of this parameter correspond to
special models discussed in the literature, \ie, $\psi(\theta=0^\circ),
\chi(\theta=-90^\circ)$, and $\eta(\theta=37.76^\circ$).

Our first analysis concerns the quantity
$r_{\nu\nu Z}\equiv\Gamma(Z'\to Z \nu \bar \nu)/\Gamma(Z' \to\lplm )$ which
has been suggested by several of the authors of Ref.~1 as a probe of the
$Z'$ couplings. However, as shown explicitly in our earlier work,
the SM background ($B$) from $ZZ$ production (with
one of the $Z$'s decaying to $\nu \bar \nu$) can be up to several orders of
magnitude larger than the $Z'\to Z\nu\bar\nu$ signal ($S$) in the
absence of sophisticated cuts. This is demonstrated in Figs.\ 1a
and 1b, where the missing $p_t$ distributions are presented for both processes
(signal and background) at the SSC
and LHC, subject to mild rapidity ($\eta$) cuts, assuming a $Z'$ mass of 1 TeV.
For the various choices of extended models, the dotted curve in these figures
must be scaled by a factor, $f$, which is model and rapidity cut dependent
and is given numerically in Table 1. Clearly, unless the
background can be precisely determined, there is very little hope of seeing
the $Z'$ signal even if both the $\eta$ and $\not p_t$ cuts are optimized.
However, we observe that the SM process $pp \to ZZ \to Z\lplm$ can be
precisely {\it {measured}} for a  fixed $p_t$ of the \lplm\ pair, subject to
the constraint that the pair invariant mass reconstructs to a $Z$. If this
set of data were then to be rescaled by the ratio of the
$Z \to \nu \bar \nu$ to $Z \to\lplm$
branching fractions, and corrected for acceptances, the background for the
$Z'$ process in question can then be determined.
As a check of this procedure, the $Z \nu \bar \nu$ data
normalization can be examined at low $p_t$ where the $Z'$ contribution is
negligible. Since the background peaks at low $p_t$ and the signal is absent
at $p_t\gsim 500$ GeV for a 1 TeV $Z'$, demanding $p_t^{min}\leq p_t \leq
500$ GeV, with $p_t^{min}$ in the 50-200 GeV range, will significantly enhance
$S/{\sqrt {B}}$. We find that a value of $p_t^{min}=200$ GeV gives the largest
$S/{\sqrt {B}}$ for both choices of the rapidity cut.

\begin{table}
\centering
\begin{tabular}{|c|c|c|c|c|} \hline\hline
Model  &\multicolumn{2}{c|} {SSC}  &\multicolumn{2}{c|} {LHC} \\ \cline{2-5}
       & 2.5    & 0.5     & 2.5   & 0.5   \\ \hline
SSM    & 2.794  & 0.664   & 0.656 & 0.213 \\
LRM    & 0.344  & 0.083   & 0.079 & 0.026 \\
ALRM   & 1.172  & 0.249   & 0.314 & 0.091 \\
$\psi$ & 0.202  & 0.047   & 0.048 & 0.015 \\
$\chi$ & 0.838  & 0.214   & 0.178 & 0.063 \\
$\eta$ & 0.090  & 0.020   & 0.022 & 0.007 \\ \hline\hline
\end{tabular}
\caption{
Values of the parameter $f$ for the SSC and LHC assuming either value
of the rapidity cut ($\eta\leq 2.5$ or $0.5$) for several different extended
electroweak models discussed in the text.}
\end{table}

To proceed further, we simulate data for both the signal and the background
within the framework of a given model, subject to the cuts above, and
determine the size of the excess sample of events obtained by integrating the
corresponding missing $p_t$ distributions.
Since the size of the background has been anticipated from the
charged lepton data as discussed above, it is then subtracted from the total
number of events.  If a statistically significant residual is obtained, it
can be attributed to a 1 TeV $Z'$. (We remind the reader
that the $Z'$ will already have been discovered and its mass will have been
determined via the charged lepton decay modes before the technique we're
discussing here is applied.)
Normalizing this excess to the number of \lplm\ events in the $Z'$ discovery
channel eliminates the systematic uncertainties due to, \eg, structure
function variations and the integrated luminosity determination. Table 2
displays the inverse of the relative error in $r_{\nu \nu Z}$, \ie,
$(\Delta r_{\nu \nu Z}/r_{\nu \nu Z})^{-1}$ (which is equivalent to
$S/\sqrt B$), obtained from this method for a number of extended gauge models.
Here, the resulting statistical and systematic errors have been combined in
quadrature.  The integrated luminosity (${\cal L}$) for the
SSC(LHC) was assumed to be 10(100) $fb^{-1}$ in obtaining these relative
errors which have magnitudes that are found to scale approximately as
$1/{\sqrt{\cal L}}$.  From the Table we see that at
least for some of the extended models, the value of $r_{\nu \nu Z}$ can be
sufficiently well determined as to provide some additional information about
the nature of the $Z'$ couplings.
It is clear from this analysis, however, that this technique
will most likely fail for larger $Z'$ masses due to the loss in statistics
necessary to observe the signal relative to the background.

\begin{table}
\centering
\begin{tabular}{|c|c|c|c|c|} \hline\hline
Model  &\multicolumn{2}{c|} {SSC}  &\multicolumn{2}{c|} {LHC} \\ \cline{2-5}
       & 2.5    & 0.5     & 2.5   & 0.5   \\ \hline
SSM    & 8.379  & 4.104   & 10.70 & 6.704 \\
LRM    & 1.032  & 0.520   & 1.289 & 0.818 \\
ALRM   & 3.515  & 1.561   & 5.123 & 2.864 \\
$\psi$ & 0.608  & 0.295   & 0.783 & 0.472 \\
$\chi$ & 2.518  & 1.342   & 2.904 & 1.983 \\
$\eta$ & 0.270  & 0.125   & 0.359 & 0.220 \\ \hline\hline
\end{tabular}
\caption{Values of the quantity
$(\Delta r_{\nu \nu Z}/r_{\nu \nu Z})^{-1}$  at the SSC and LHC
assuming  either choice of the rapidity cut.}
\end{table}

A second approach to obtaining information about the $Z'$ coupling to
$\nu \bar \nu$ is to examine the production of a high $p_t$ jet in association
with both the $Z'$ as well as the SM $Z$. The essential idea is an extension of
the monojet analysis by the CDF Collaboration{\cite {5}} from which they
obtained a constraint on the number of neutrinos. We consider the following
four processes: $pp \to Z+jet , Z'+jet \to\lplm +jet, \nu \bar \nu +jet$ at
both the SSC and LHC as functions of the jet transverse momentum, assuming
$|\eta_j| <0.5$ or 2.5. (In practice, the $\eta_j$ cut is chosen to be 2.5
since the lower value only results in decreased statistics.) The two
processes which involve charged leptons in the final state can be directly
measured, with extremely high statistics in the case of the SM $Z$, and are
easily separable
by demanding the dilepton pairs lie in invariant mass bins associated with the
$Z$ or the $Z'$. (We again assume that the $Z'$ mass is already
well determined from the discovery channel; for purposes of demonstration
we take its value to be 1 TeV.) From the number of observed SM $Z$ induced
$\lplm j$ events we can deduce the anticipated number of SM
$Z$ monojet events, which we treat as a background, through again
simply scaling by the well measured ratio of
$Z \to \nu \bar \nu$ to $Z \to\lplm$ branching fractions and correcting
for acceptance differences. Of course, the monojet rate also obtains a
contribution from the $Z'+jet$ intermediate state, which is proportional to
the $Z' \to \nu \bar \nu$ branching fraction, and may appear as a signal
over the SM $Z$ background. For a given extended electroweak model, we
simulate both the signal and the
backgrounds for a minimum jet $p_t$ in the $50-200$ GeV range allowing for
statistical fluctuations in the number of events. We subtract from the total
event sample, after integrating over $\eta$ and $p_t$, the anticipated
number of SM events, obtained through the rescaling technique mentioned above
and normalize any potential excess by the number of $Z'$ induced
$\lplm j$ events. If we assume that other new physics sources of monojets are
not present, this normalized excess of events is directly proportional to the
quantity $R_{\nu}=\Gamma(Z' \to \nu \bar \nu)/\Gamma(Z' \to\lplm)$, which
is quite sensitive to $Z'$ coupling variations.  (Assuming
that invisible decays of the $Z'$ arise solely from its decay into the 3
generations of SM left-handed $\nu$'s.) In taking the ratio above we again
have eliminated potentially large systematic errors arising from variations in
the parton distributions and integrated luminosity uncertainties.

What range of values can we expect for the ratio $R_{\nu}$? In extended models
where the new generator
which couples to the $Z'$ commutes with the SM $SU(2)_L$ generators, we find
that $R_{\nu}$ can be expressed as $R_{\nu}=3c_L^2/(c_L^2+c_R^2)$, with
$c_{L(R)}$ being the left(right)-handed coupling of the charged lepton to the
$Z'$. Clearly, in this case, we must have $0 \leq R_{\nu} \leq 3$,
independently of the specific choice of model. When $R_{\nu}=0(3)$, the
$Z'\lplm$ coupling is purely right(left)-handed. This `commuting'
condition
is readily satisfied by the LRM, ALRM, and ER5M choices but {\it not} for the
SSM case since the $SU(2)_L$ generators do not commute amongst themselves.
Figs.\ 2a-c show the anticipated values of $R_{\nu}$ in the ER5M case,
and the associated errors which
follow from this analysis, as functions of $\theta$ at both the SSC and LHC
with
various cuts on the jet transverse momentum. We see immediately that ($i$)
$R_{\nu}$ is quite sensitive to the value of $\theta$ (thus demonstrating
its usefulness as a model discriminator), ($ii$)
a stricter $p_t$ cut reduces the size of the anticipated errors, and ($iii$)
the same cuts applied to the data from the SSC will yield smaller errors than
that similarly obtained at the LHC even when the
factor of 10 difference in integrated luminosity is accounted for.
In Fig.\ 3a, we see the corresponding results for the LRM as a function
of $\kappa\equiv g_R/g_L$, the ratio
of the $SU(2)_R$ to $SU(2)_L$ gauge couplings. For large values of $\kappa$ we
note that the couplings of the charged leptons become purely right-handed. As
in the case of the parameter
$\theta$ in the ER5M, $R_{\nu}$ is quite sensitive to the value of
$\kappa$, particularly when it lies in its natural range, \ie, $0.55 \leq
\kappa \leq 1$, anticipated by GUT analyses{\cite {6}}. Fig.\ 3b displays a
comparison of the values of $R_{\nu}$ obtained for the ALRM, SSM, LRM with
$\kappa=1$, and ER5M as a function of $\theta$. It is clear from the figure
that $R_{\nu}$ can be determined with reasonable precision and may be an
extremely useful discriminator of extended gauge models provided the
assumptions we have made above are valid. The presence of monojets from
other new physics sources, such as SUSY, will render this approach
extremely difficult if not impossible. As in the previous analysis we observe
that for substantially larger $Z'$ masses there will be insufficient
statistics to overcome the large SM backgrounds.

In the analyses above we have considered two possible ways of gaining
information on the couplings of a $Z'$ to neutrinos in the face of large SM
backgrounds. Rather than relying solely on theoretical calculations, which
suffer from many uncertainties, we propose instead to use data from SSC and/or
LHC experiments themselves to determine the size of these backgrounds. In
following this route and always employing ratios of numbers of events,
many of the systematic uncertainties are found to cancel. For the
first possibility examined, the case of three-body $Z'$ decays, we found that
it may be possible to resurrect the quantity $r_{\nu \nu Z}$ as an model
discriminator provided strong $p_t$ cuts and invariant mass binning can be
applied to the data without appreciable loss in rate. For many extended
electroweak models, this
technique was found to work reasonably well for a $Z'$ in the 1 TeV range, but
for significantly larger masses, there is a dramatic loss in
statistical power.
A second technique that employs potential excesses in monojet events as a
signal for $Z' \to \nu \bar \nu$, is extremely dependent upon several
assumptions, in particular, that no other non-SM sources of monojets, such as
SUSY, exist. If such sources are absent (or possibly well determined by other
means), it was found that the ratio of the
partial widths of the $Z'$ to $\nu \bar \nu$ and \lplm\ can be reasonably
well determined for $Z'$ masses in the 1 TeV range.

\vspace{0.3cm}

\noindent{\bf Acknowledgements}

This work has been supported by the U.S. Department of Energy, Division of
High Energy Physics under contract W-31-109-ENG-38.

\vspace{0.3cm}
%
\def\MPL #1 #2 #3 {Mod.~Phys.~Lett.~{\bf#1},\ #2 (#3)}
\def\NPB #1 #2 #3 {Nucl.~Phys.~{\bf#1},\ #2 (#3)}
\def\PLB #1 #2 #3 {Phys.~Lett.~{\bf#1},\ #2 (#3)}
\def\PR #1 #2 #3 {Phys.~Rep.~{\bf#1},\ #2 (#3)}
\def\PRD #1 #2 #3 {Phys.~Rev.~{\bf#1},\ #2 (#3)}
\def\PRL #1 #2 #3 {Phys.~Rev.~Lett.~{\bf#1},\ #2 (#3)}
\def\RMP #1 #2 #3 {Rev.~Mod.~Phys.~{\bf#1},\ #2 (#3)}
\def\ZP #1 #2 #3 {Z.~Phys.~{\bf#1},\ #2 (#3)}
\def\IJMP #1 #2 #3 {Int.~J.~Mod.~Phys.~{\bf#1},\ #2 (#3)}
\bibliographystyle{unsrt}

\newpage

\noindent
Fig.~1: Missing $p_t$ distribution for the background process, $pp \to 2Z \to
Z \nu \bar \nu$, at the the SSC(solid) and the LHC(dash-dotted) as well as for
the signal(dotted), $pp \to Z' \to Z \nu \bar \nu$, assuming an $\eta$ cut of
(a)2.5 or (b)0.5.

\medskip

\noindent
Fig.~2: The ratio $R_{\nu}$ as a function of the parameter $\theta$ in the
ER5M assuming a $Z'$ mass of 1 TeV with $|\eta_j|<2.5$ at (a)the SSC with
$p_t>200$ GeV, or (b) with $p_t>50$ GeV. In either case, the errors reflect
the combined statistical and systematic errors associated with an integrated
luminosity of $10 fb^{-1}$. (c) Same as (a) but for the LHC with an integrated
luminosity of $100 fb^{-1}$.
\medskip

\noindent
Fig.~3: (a)Same as Fig.~2a but for the LRM as a function of the parameter
$\kappa\equiv g_R/g_L$.
(b) A comparison of the values for $R_{\nu}$ as in Fig.~2a
for the ALRM, SSM, LRM with $\kappa=1$, and ER5M as a function of $\theta$.

\end{document}